\documentclass[showpacs,preprintnumbers,amsmath,amssymb,10pt]{revtex4}
\usepackage{amsmath}
\usepackage{graphicx}
\usepackage{subfig}
\usepackage{hyperref}
\usepackage{amssymb}
\usepackage[english]{babel}
\usepackage{epsfig}
\usepackage{wasysym}

\begin{document}
\title{Reconstruction, Thermodynamics and Stability of $\Lambda$CDM Model in $f(T,\mathcal{T})$ Gravity \\}
\author{Ednaldo L. B. Junior$^{(a,e)}$}\email{ednaldobarrosjr@gmail.com}
\author{Manuel E. Rodrigues$^{(a,b)}$}\email{esialg@gmail.com} 
\author{Ines G. Salako$^{(c)}$}\email{inessalako@gmail.com}
\author{Mahouton J. S. Houndjo$^{(c,d)}$}\email{sthoundjo@yahoo.fr}

\affiliation{$^{(a)}$ Faculdade de F\'{\i}sica, PPGF, Universidade Federal do Par\'{a}, 66075-110, Bel\'{e}m, Par\'{a}, Brazil}

\affiliation{$^{(b)}$ Faculdade de Ci\^{e}ncias Exatas e Tecnologia, Universidade Federal do Par\'{a}\\
Campus Universit\'{a}rio de Abaetetuba, CEP 68440-000, Abaetetuba, Par\'{a}, Brazil}

\affiliation{$^{(c)}$ Institut de Math\'{e}matiques et de Sciences Physiques (IMSP),  
01 BP 613, Porto-Novo, B\'{e}nin}

\affiliation{$^{(d)}$ Facult\'{e} des Sciences et Techniques de Natitingou - Universit\'{e} de Parakou - B\'{e}nin}

\affiliation{$^{(e)}$ Faculdade de Engenharia da Computa\c{c}\~{a}o, Universidade Federal do Par\'{a}, Campus Universitário de Tucuru\'{\i}, CEP: 68464-000, Tucuru\'{\i}, Par\'{a}, Brazil}
\begin{abstract}
We reconstruct the $\Lambda$CDM model for  $f(T,\mathcal{T})$ Theory, where $T$ is the torsion scalar and $\mathcal{T}$ the trace of the energy-momentum tensor. The result shows that the action of $\Lambda$CDM is a combination of  a linear term, a constant ($-2\Lambda$) and a non-linear term given by the product  $\sqrt{-T}F_g\left[(T^{1/3}/16\pi G)\left(16\pi G\mathcal{T}+T+8\Lambda\right)\right]$, with $F_g$ being a generic function. We show that to maintain conservation of energy-momentum tensor should impose that $F_g[y]$ must be linear on the trace $\mathcal{T}$. This reconstruction decays in the  $f(T)$ Theory for $F_g\equiv Q$, with $Q$ a constant. Our reconstruction describes the cosmological eras to the present time. The model present stability within the geometric and matter perturbations for the choice $F_g=y$, where $y=(T^{1/3}/16\pi G)\left(16\pi G\mathcal{T}+T+8\Lambda\right)$, except for geometric part to de Sitter model. We impose the first and second laws of thermodynamics to the $\Lambda$CDM and find the condition where they are satisfied, that is,  $T_A,G_{eff}>0$, however where this is not possible for cases where we choose, leading to a breakdown of positive entropy and Misner-Sharp energy.
\end{abstract}

\maketitle

\section{Introduction}
\label{sec1}

The description of the gravitational interaction can be done in different ways, the main and best known is that using Riemannian geometry as a tool for the formulation of this description. This is known as the Theory of General Relativity (GR) of Einstein \cite{wald}.
\par 
The Riemannian geometry is based on the differential geometry which deals with the so-called space-time as a differentiable manifold of dimension four, where we have only the effect of the curvature. There are other geometries that generalize this idea. A space-time may have curvature, but also torsion in their structure. There are two fundamental concepts in differential geometry. The formulation of this type of geometry has been undertaken and gravitation is known as Einstein-Cartan geometry \cite{EC}. In this theory the gravitational interaction is described by both curvature and torsion of space-time, where the torsion is commonly attributed to the inclusion of spin, through fractional spin fields. A very particular case of this theory is when we take the identically zero curvature, and then only have a space-time with torsion. This type of geometry is known as the Weitzenbock geometry \cite{ortin, weitzenbock}, where the torsion describes the gravitational interaction. Various analysis can be performed 
in this type of geometry, which is proven to be dynamically equivalent to GR \cite{maluf1}. In this context, recently a new formulation is started up, and generalizes the call Teleparallel Theory (TT), of the space-time where the gravitational interaction is described solely by the torsion.
\par 
Through the standard Big Bang \cite{mukhanov} theory, the $\Lambda$CDM models describe very well the evolution of our universe, in a Riemannian geometry within the GR. There are several open issues, but the main one today is the so-called dark energy. In order to make the cosmological WMAP data fitting with the theory, it is necessary to introduce an exotic component in the equations of GR, i.e, the dark energy. This can be modelled as a perfect fluid with the equation of state $ p_{DE} = \omega_{DE} \rho_{DE}$, where the values of $\omega_{DE}$ must be very close to the $-1$ today. An alternative to that data, being consistent with the theory, is the modification of the geometry. A good review can be seen in \cite{bamba}. One of the first general possibilities is the well known $f(R)$ Theory \cite{faraoni}, where $R$ is the scalar curvature, obtained through the double contraction of the Riemann tensor indices. This theory places the analytic function $f(R)$ in the action, where the GR can be reobtained in 
a certain limit, such as from $f(R)= a_1R+ a_2 R^2 $, with $ a_2 \rightarrow 0 $ one gets again GR. This theory has proved effective in simulating the evolution of our universe, in various epochs. Other possibilities have arisen through the generalisation of the GR, by changing the action. One such change is the case of  $f(R, \mathcal{T})$ Theory, where $\mathcal{T}$ is the trace of the energy-momentum tensor. In this case, the matter content should be taken into account as  having a kind of interaction with the geometry.
\par 
A direct analogy could be made between theories with only the curvature and the torsion only. As the $f(R)$ theory is a generalization of the GR, it is logic to also think of a generalization of the TT, where the analogous object to the riemannian scalar curvature, is the torsion scalar $T$, obtained from contractions  between the torsion and contorsion tensors. A change in the action of TT is made considering an analytic function $f(T)$ which depends on the torsion scalar. That was first thought of a theory arising from the Born-Infeld action \cite{ferraro1}. Then, several studies have shown the great accordance of this theory with the most varied approaches in Gravitation and Cosmology \cite{fT}.
\par 
Another recent proposal is to consider not only the torsion scalar in the action, but also the trace of the energy-momentum tensor, as an analogy to the $ f(R, \mathcal{T})$ Theory. This theory, called $ f(T, \mathcal{T})$ has been formulated recently \cite{saridakis1}, still requiring verification of compatibility with the cosmological data and the physical requirements for a good cosmological theory. That is the reason  to check  how should be the functional form of the action of this theory, such that the $\Lambda$CDM model is valid. For this we use the scheme the reconstruction method for a modified gravity \cite{odintsov1}. In addition, we make a stability analysis for the $\Lambda$CDM model.
\par 
It also exists a wide interest in studying the thermodynamics of our universe. Various calculus have been done for guaranteeing the system to obey the first and second laws of thermodynamics. In the GR, it has been shown that the first law of thermodynamics can be shown as $dE=TdS+WdV$ \cite{akbar}. This law can also be represented in the modified versions of gravity, but with an additive content of entropy production, for a non-equilibrium description \cite{bamba2}. We have done here an analysis which yields the conditions for satisfying the classical thermodynamics laws. We adopt the units as $k_{B}=c=\hbar=1$ and the Newton constant $G^{-1/2}=M_{Plank}=1.2\times 10^{19}GeV$ \cite{kolb}.
\par
The paper is organized as follows. In section \ref{sec2} we make a brief description of $ f(T)$ and $f(T, \mathcal{T})$ theories, with the main elements and definitions necessary for their formulation. In section \ref{sec2.1} we have analysed the conservation of energy-momentum tensor, {\bf which results in strong constraints to the functional form of the  action of the theory, leading to function $f(T, \mathcal{T})$ with a linear dependence on the trace $\mathcal{T}$}. In the \ref{sec3} section, we use the reconstruction method for the actions to obtain the $\Lambda$CDM model for $f(T, \mathcal{T})$ theory. {\bf The result shows that the action of $\Lambda$CDM is a combination of  a linear term, a constant ($-2\Lambda$) and a non-linear term given by the product  $\sqrt{-T}F_g\left[(T^{1/3}/16\pi G)\left(16\pi G\mathcal{T}+T+8\Lambda\right)\right]$, with $F_g$ being a generic function}. In section \ref{sec4} we make the stability analysis of the studied model. {\bf The model present stability within the geometric and matter perturbations for the choice $F_g=y$, where $y=(T^{1/3}/16\pi G)\left(16\pi G\mathcal{T}+T+8\Lambda\right)$, except for geometric part to de Sitter model}. We do a thermodynamic analysis for $f(T,\mathcal{T})$ theory in section \ref{sec5}. {\bf We impose the first and second laws of thermodynamics to the $\Lambda$CDM and find the condition where they are satisfied, that is,  $T_A,G_{eff}>0$, however where this is not possible for cases where we choose, leading to a breakdown of positive entropy and Misner-Sharp energy}. In section \ref{sec5.1} we have established the conservation of energy-momentum tensor to previous results, {\bf showing an inconsistency for this approach}. We make our final considerations in section \ref{sec6}.

\section{$f(T)$ and $f(T,\mathcal{T})$ gravities}
\label{sec2}
In this section we will see the basic preliminary concepts for the reconstruction of theories $f(T)$ and $f(T, \mathcal{T})$ theories of gravity.
\par
In theory $f(T)$, the geometry is determined solely by the matrices that transform the metric of space-time into the Minkowski metric. To begin, we define the space-time as a differentiable manifold in which only the torsion is non-zero, that is, the curvature is identically zero, then all Riemann tensor components  are zero.
\par
Now we define the line element as
\begin{eqnarray}
dS^2=g_{\mu\nu}dx^{\mu}dx^{\nu}\,.
\end{eqnarray} 
Taking into account that we can define 1-forms in the co-tangent space of the manifold, and introduce Lorentz symmetry in the line element, we can re-write the line element as
\begin{eqnarray}
dS^2=g_{\mu\nu}dx^{\mu}dx^{\nu}=\eta_{ab}\theta^a\theta^b\,,\label{metric}
\end{eqnarray}
where $\theta^a=e^a_{\;\;\mu}dx^{\mu}$ are 1-forms, index  $a=0,...,3$ and $[\eta_{ab}]=diag[1,-1,-1,-1]$ is the Minkowski metric. Here the Latin indices are related to the co-tangent space and the Greeks ones to the space-time.   By the way of writting the line element  (\ref{metric}), we can establish the following relations $\eta_{ab}=e_{a}^{\;\;\mu}e_{b}^{\;\;\nu}g_{\mu\nu}$, $g_{\mu\nu}=e^{a}_{\;\;\mu}e^{b}_{\;\;\nu}\eta_{ab}$, $e^{a}_{\;\;\mu}e_{a}^{\;\;\nu}=\delta^{\nu}_{\mu}$, $e_{a}^{\;\;\nu}e^{b}_{\;\;\nu}=\delta^{b}_{a}$, where $e_{a}^{\;\;\mu}$ is the inverse of matrix called tetrad  $e^{a}_{\;\;\mu}$. 
\par 
The connection is chosen such that all the Riemann tensor components are identically zero, and one has the Weitzenbock connection \cite{saridakis2}
\begin{eqnarray}
\Gamma^{\alpha}_{\;\;\nu\mu}=e_{a}^{\;\;\alpha}\partial_{\nu}e^{a}_{\mu}\label{connection}\,.
\end{eqnarray}  
We can now define a tensor which gives sense of torsion to the space-time 
\begin{eqnarray}
T^{\alpha}_{\;\;\mu\nu}=\Gamma^{\alpha}_{\;\;\nu\mu}-\Gamma^{\alpha}_{\;\;\mu\nu}=e_{a}^{\;\;\alpha}\left(\partial_{\mu}e^{a}_{\;\;\nu}-\partial_{\nu}e^{a}_{\;\;\mu}\right)\label{torsion}\,.
\end{eqnarray}
Through the components of the torsion tensor, we can define the contortion tensor components and the tensor $S_{\alpha}^{\;\;\mu\nu}$ 
\begin{eqnarray}
&&K^{\mu\nu}_{\;\;\;\;\alpha}=-\frac{1}{2}\left(T^{\mu\nu}_{\;\;\;\;\alpha}-T^{\nu\mu}_{\;\;\;\;\alpha}-T_{\alpha}^{\;\;\mu\nu}\right)\,,\label{contorsion}\\
&&S_{\alpha}^{\;\;\mu\nu}=\frac{1}{2}\left(K^{\mu\nu}_{\;\;\;\;\alpha}+\delta^{\mu}_{\alpha}T^{\sigma\nu}_{\;\;\;\;\sigma}-\delta^{\nu}_{\alpha}T^{\sigma\mu}_{\;\;\;\;\sigma}\right)\label{S}\,.
\end{eqnarray}
We can also define the analogous object to the scalar curvature in GR, the torsion scalar 
\begin{eqnarray}
T=T^{\alpha}_{\;\;\mu\nu}S_{\alpha}^{\;\;\mu\nu}=\frac{1}{4}T^{\mu\nu\sigma}T_{\mu\nu\sigma}+\frac{1}{2}T^{\mu\nu\sigma}T_{\sigma\nu\mu}-T_{\sigma}^{\;\;\sigma\mu}T^{\nu}_{\;\;\nu\mu}\label{storsion}\,.
\end{eqnarray}
It is this object that plays the curvature scalar role in GR, and that should also form the action of $f(T)$ theory. The action of $f(T)$ theory is constructed in a way that we have a linear term in the torsion scalar, another containing the correction term  to the TT and another term related to the material content. Then we write the action as
\begin{eqnarray}
S= \frac{1}{16\pi G }\int d^4x e
\left[T+f(T)+\mathcal{L}_m\right],\label{actionfT},
\end{eqnarray}
with $e = det[e^{a}_{\;\;\mu}] = \sqrt{-g}=\sqrt{det[g_{\mu\nu}]}$, $G$ the Newton's constant, and setting $c$ to unity. 
In $ f(T)$ theory, tetrads are dynamic fields, then doing the functional variation of the action (\ref{actionfT}) in relation to them, one gets the  following equations of motion
\begin{eqnarray}
\label{eqm1}
&&\left(1+f_T\right)\,\left[e^{-1}\partial_{\mu}(ee_a^{\;\;\sigma}S_{\sigma}^{\;\;\nu\mu
}) -e_{a}^{\;\;\lambda}T^{\sigma}_{\;\;\mu\lambda}S_{\sigma}^{\;\;\mu\nu}\right]
 + e_a^{\;\;\sigma}S_{\sigma}^{\;\;\nu\mu}\partial_{\mu}{T} f_{TT}+\frac{1}{4} e_ {a}^{\;\;\nu}[T+f] = 4\pi Ge_{a}^{\sigma}\Theta_{\sigma}^{\;\;\nu},
\end{eqnarray}
where we use the nomenclature $ f_T = \partial f/\partial T $ and $ f_{TT} = \partial^{2}f/\partial T^{2}$, $\Theta_{\sigma}^{\;\;\nu}$ represents the components of the matter energy-momentum tensor.
\par
Let's take the example of the Friedmann-Lema\^itre-Robertson-Walker (FLRW) universe with flat spatial section
\begin{equation}
ds^2= dt^2-a^2(t)\left(dx^2+dy^2+dz^2\right) \,,\label{FLRW}
\end{equation}
where $a(t)$ is the scale factor. The Hubble parameter is given by $H(t)=\dot{a}(t)/a(t)$. Now we specify our choice of tetrads as
\begin{equation}
\label{tetrad1}
[e^{a}_{\;\;\mu}]=diag[1,a(t),a(t),a(t)].
\end{equation}
With this choice, we can represent the line element (\ref{metric}) through a set of 1-forms $\big[\theta^0=dt,$ $\theta^1=a(t)dx,\theta^2=a(t)dy,\theta^3=a(t)dz\big]$.
Thus, the equations of motion (\ref{eqm1}) for $f(T)$ theory, taking the material content as a perfect fluid $\Theta_{\mu}^{\;\;\nu}=diag[\rho_{mt},-p_{mt},-p_{mt},-p_{mt}]$, are given as follows
\begin{eqnarray}\label{eqm2-1}
&&H^2= \frac{8\pi G}{3}\rho_{mt}
-\frac{f}{6}-2H^2f_T\\
\label{eqm2-2}
&&\dot{H}=-\frac{4\pi G(\rho_{mt}+p_{mt})}{1+f_T-12H^2f_{TT}},
\end{eqnarray}
The subscribed $\rho_{mt}$ and $p_{mt}$ means the density and pressure of the total matter in the universe, we consider here only the components of the baryonic matter as $\{\rho_m, p_m\}$ and radiation as $\{\rho_r, p_r \}$. 
\par 
Here, the torsion scalar is obtained by the definitions (\ref{torsion})-(\ref{storsion}) with  (\ref{tetrad1}), resulting in
\begin{eqnarray}
\label{storsion2}
T=-6H^2,
\end{eqnarray}
We now see clearly that the equations of motion of $f(T)$ theory are identical to that of the  GR, and the equations of Friedmann (flat spatial section), when the nonlinear term are zero, i.e making $f(T )=f_T=f_{TT}=0$ in (\ref{eqm2-1})-(\ref{eqm2-2}).\par
Now we can present the most recent generalization of the $f(T)$ theory. Following the analogy of the generalization of the $f(R)$ theory to $f(R, \mathcal{T})$, where $\mathcal{T}$ is the trace of the energy-momentum tensor. Here, we can also introduce in the action an analytic function that depends not only on the torsion scalar, but also on the trace $\mathcal{T}$. For such a theory, the action is then given by
\begin{equation}
S= \frac{1}{16\,\pi\,G}\,\int d^{4}x\,e\,\left[T+f(T,\mathcal{T})+16¨\pi G\mathcal{L}_{m}\right]\,,\label{action2}
\end{equation}
where $f(T,\mathcal{T})$ is an arbitrary analytical function of the torsion scalar $T$ and of the trace $\mathcal{T}$ of the matter energy-momentum tensor  $\Theta_{\mu}^{\;\;\nu}$, and  $\mathcal{L}_{m}$ is the matter Lagrangian density.\par 
Here we must consider that the Lagrangian density $ \mathcal{L}_{m}$ depends only on tetrads and not on its derivatives. For the energy-momentum tensor of a perfect fluid we have the following trace
\begin{eqnarray}
\mathcal{T}=\Theta_{\mu}^{\;\;\mu}=\rho_{mt}-3p_{mt}\,.\label{trace}
\end{eqnarray}
We can then make the functional variation of the action (\ref{action2}) with respect to the tetrads, resulting in the following equations of motion \cite{saridakis1}
\begin{eqnarray}
&&\left(1+f_{T}\right) \left[e^{-1} \partial_{\mu}{(e
e^{\;\;\alpha}_{a}
S_{\alpha}^{\;\;\sigma \mu})}-e^{\;\;\alpha}_{a} T^{\mu}_{\;\;\nu \alpha} S_{\mu}^{\;\;\nu
\sigma}\right] 
+\left(f_{TT} \partial_{\mu}{T}+f_{T\mathcal{T}} \partial_{\mu}
\mathcal{T}\right) e^{\;\;\alpha}_{a} S_{\alpha}^{\;\;\sigma \mu}
\nonumber \\
&&+ e_{a}^{\;\;\sigma}
\left(\frac{f+T}{4}\right)  -f_{\mathcal{T}} \left(\frac{e^{\;\;\alpha}_{a} \Theta_{\alpha}^{\;\;\sigma}+p_m e^{\;\;\sigma}_{a}}{2}\right)=4\pi G e^{\;\;\alpha}_{a}
\Theta_{\alpha}^{\;\;\sigma},\label{eqm}
\label{eqm3}
\end{eqnarray}
where $f_{\mathcal{T}}=\partial{f}/\partial{\mathcal{T}}$ and
$f_{T\mathcal{T
}}=\partial^2{f}/\partial{T} \partial{\mathcal{T}}$. Here it is evident that the particular case where the function  $f$ depend only on the torsion scalar $T$, i.e, $f\equiv f(T)$\footnote{ Then, for $f_{\mathcal{T}}=f_{T\mathcal{T}}\equiv 0$.}, the equation (\ref{eqm3}) reduces to the equation of motion of $f(T)$ theory in (\ref{eqm1}).
\par 
Taking again the matter content as a perfect fluid and choosing the diagonal tetrads in  (\ref{tetrad1}), the equations of motion of  $f(T,\mathcal{T})$ theory for a flat FLRW universe, are given by  
\begin{eqnarray}
&&3H^{2}=8\pi G \rho_{mt}-\frac{1}{2}\left(f+12 H^{2} f_{T}\right)+f_{\mathcal{T}} \left(\rho_{mt}+p_{mt}\right)
\label{eqm4}\,,\\
 &&\dot{H}=-4\pi G \left(\rho_{mt}+p_{mt}\right)-\dot{H} \left(f_{T}-12 H^{2}
f_{TT}\right)    -H \left(\dot{\rho}_{mt}-3\,\dot{p}_{mt}\right)
f_{T\mathcal{T
}}-f_{\mathcal{T}} \left(\frac{\rho_{mt}+p_{mt}}{2} \right). 
\end{eqnarray}
In this paper, in the next section, we propose to reconstruct the  $\Lambda$CDM model and study the stability of the de Sitter and power law solutions.

\section{Conservation laws to $f(T,\mathcal{T})$ theory}\label{sec2.1}

This section is devoted to establishing a conservation law for the $f(T,\mathcal{T})$ theory. For this, we must describe the equations of motion in a covariant form. Let us multiply the equation of motion (\ref{eqm}) by $g_{\mu\sigma}e^{a}_{\;\;\nu}$ (to have the indexes $\mu$ and $\nu$ the free in the end) and using the identity $g_{\mu\sigma}e^{a}_{\;\;\nu}[e^{-1}\partial_{\lambda}(ee_{a}^{\;\;\alpha}S_{\alpha}^{\;\;\sigma\lambda})-e_{a}^{\;\;\alpha}T^{\lambda}_{\;\;\gamma\alpha}S_{\lambda}^{\;\;\gamma\sigma}]=(1/2)\left[G_{\mu\nu}-(1/2)g_{\mu\nu}T\right]$, where $G_{\mu\nu}$ is the Einstein tensor, we can rewrite our equation of motion as
\begin{eqnarray}
\frac{1}{2}(1+f_T)\left(G_{\mu\nu}-\frac{1}{2}g_{\mu\nu}T\right)+S_{\nu\mu}^{\;\;\;\;\lambda}\left(f_{TT}\partial_{\lambda}T+f_{T\mathcal{T}}\partial_{\lambda}\mathcal{T}\right)+\frac{1}{4}g_{\mu\nu}\left(f+T\right)-\frac{1}{2}f_{\mathcal{T}}\left(\Theta_{\nu\mu}+g_{\mu\nu}p_{mt}\right)=4\pi G\Theta_{\nu\mu}\label{eqmm}\,.
\end{eqnarray}
Now we take the divergence $\nabla^{\mu}$ all two sides and isolate $\nabla_{\mu}\Theta_{\nu}^{\;\;\mu}$, taking into account that $\nabla_{\mu}G_{\nu}^{\;\;\mu}\equiv 0$, what gives us
\begin{eqnarray}
&&\nabla_{\mu}\Theta_{\nu}^{\;\;\mu}=\frac{1}{\left(4\pi G+(1/2)f_{\mathcal{T}}\right)}\Bigg\{\left(f_{TT}\partial_{\sigma}T+f_{T\mathcal{T}}\partial_{\sigma}\mathcal{T}\right)e^{a}_{\;\;\nu}[e^{-1}\partial_{\lambda}(ee_{a}^{\;\;\alpha}S_{\alpha}^{\;\;\sigma\lambda})-e_{a}^{\;\;\alpha}T^{\lambda}_{\;\;\gamma\alpha}S_{\lambda}^{\;\;\gamma\sigma}]-\frac{1}{4}(1+f_T)\partial_{\nu}T\nonumber\\
&&+\nabla_{\mu}S_{\nu}^{\;\;\mu\lambda}\left(f_{TT}\partial_{\lambda}T+f_{T\mathcal{T}}\partial_{\lambda}\mathcal{T}\right)+S_{\nu}^{\;\;\mu\lambda}\Big(f_{TTT}\partial_{\mu}T\partial_{\lambda}T+f_{TT\mathcal{T}}\partial_{\mu}\mathcal{T}\partial_{\lambda}T+f_{TT}\nabla_{\mu}\partial_{\lambda}T+f_{T\mathcal{T}T}\partial_{\mu}T\partial_{\lambda}\mathcal{T}\nonumber\\
&&+f_{T\mathcal{T}\mathcal{T}}\partial_{\mu}\mathcal{T}\partial_{\lambda}\mathcal{T}+f_{T\mathcal{T}}\nabla_{\mu}\partial_{\lambda}\mathcal{T}\Big)+\frac{1}{4}\left(f_T\partial_{\nu}T+f_{\mathcal{T}}\partial_{\nu}\mathcal{T}+\partial_{\nu}T\right)-\frac{1}{2}\left(f_{\mathcal{T}T}\partial_{\mu}T+f_{\mathcal{T}\mathcal{T}}\partial_{\mu}\mathcal{T}\right)\left(\Theta_{\nu}^{\;\;\mu}+\delta^{\mu}_{\nu}p_{mt}\right)\nonumber\\
&&-\frac{1}{2}f_{\mathcal{T}}\partial_{\nu}p_{mt}\Bigg\}\label{eqnonconserv}\,.
\end{eqnarray}

This equation gives us two constraints, one for $\nu=0$ and other for $\nu=1,2,3$, as follows
\begin{eqnarray}
\dot{\rho}_{mt}+3H(\rho_{mt}+p_{mt})&=&\frac{1}{4}\Bigg\{2(\rho_{mt}+p_{mt})\left(f_{T\mathcal{T}}\frac{dT}{dt}+f_{\mathcal{T}\mathcal{T}}\frac{d\mathcal{T}}{dt}\right)-f_{\mathcal{T}}\frac{d\mathcal{T}}{dt}+2f_{T}\frac{dp_{mt}}{dt}\Bigg\}\label{eqnonconser1}\,,\\
0&=& \frac{p_{mt}}{2}\left[f_{T\mathcal{T}}\frac{dT}{dt}+f_{\mathcal{T}\mathcal{T}}\frac{d\mathcal{T}}{dt}\right]\label{eqnonconserv2}\,.
\end{eqnarray}

The second constraint results in
\begin{eqnarray}
\frac{d}{dt}[f_{\mathcal{T}}]=0\label{eqnonconserv2.1}\,,
\end{eqnarray}
fixing $f(T,\mathcal{T})$ as a linear function of the trace $\mathcal{T}$ or a constant, which is in agreement with the results obtained recently in \cite{fRT2015} for the analogous $f(R,\mathcal{T})$ Theory. Already through the first constraint, if we are to maintain the conservation of energy-momentum tensor, we must impose $\dot{\rho}_{mt}+3H(\rho_{mt}+p_{mt})\equiv 0$, as usually obtained in the $\Lambda$CDM model. This provides us 
\begin{eqnarray}
2(\rho_{mt}+p_{mt})\left(f_{T\mathcal{T}}\frac{dT}{dt}+f_{\mathcal{T}\mathcal{T}}\frac{d\mathcal{T}}{dt}\right)-f_{\mathcal{T}}\frac{d\mathcal{T}}{dt}+2f_{T}\frac{dp_{mt}}{dt}=0\label{eqnonconserv1.1}\,.
\end{eqnarray}

When we do the same analysis in $f(R,\mathcal{T})$ Gravity, we also have a restriction on the functional form of the action \cite{alvaro}.
\par 
We will guard these results and reconstruct the $\Lambda$CDM model taking into account these two constraint.

\section{Reconstrction of $\Lambda$CDM model in $f(T,\mathcal{T})$ theory}\label{sec3}


In this section we use the reconstruction method, through a particular model,  for obtaining what should be the functional form of the function $f(T,\mathcal{T})$. This method basically consist in choosing a model consistent with the cosmological data, and use the imposition of the characteristic equation for this specific model, must be satisfied at any time of the evolution of our universe. So with an imposition, we integrate the equation of motion in order to make the model being valid. This results in a functional form of fixed function that makes up the action of the theory. The use of this method in some cases of modified gravity can be seen in \cite{odintsov1,karami}.
\par 
Now, for reconstructing the  $\Lambda$CDM model,  it is necessary to impose an equation of motion of this model
\begin{eqnarray}
3H^2=8\pi G\rho_{mt}+\Lambda\label{LCDM}\,,
\end{eqnarray}
where $\Lambda$ is the cosmological constant. In this model, the matter is described through a perfect fluid formulation, where the pressure satisfies to the following equation of state $p_{mt}=\omega_m\rho_m+\omega_r\rho_r=(1/3)\rho_r$, with $\rho_m$ and $\rho_r$ being the matter and the radiation densities. The trace of the energy-momentum tensor  (\ref{trace}) is 
\begin{eqnarray}
\mathcal{T}=\rho_{mt}-3p_{mt}=(1-3\omega_m)\rho_m+(1-3\omega_r)\rho_r=\rho_m\label{trace1.0}\,.
\end{eqnarray}

By inverting the relation (\ref{trace1.0}) for the energy matter density in terms of the trace and using (\ref{storsion2}), imposing the  $\Lambda$CDM model from the equation (\ref{LCDM})\footnote{We can extract of the imposition of the $\Lambda$CDM model the following relationship $\rho_r=-\left\{(1/8\pi G)[(T/2)+\Lambda]+\mathcal{T}\right\}$.}, we can substitute in the equation of motion  (\ref{eqm4}) of the  $f(T,\mathcal{T})$ theory, yielding 
\begin{eqnarray}
\Lambda=-\frac{1}{2}(f-2Tf_T)-\frac{1}{3}f_{\mathcal{T}}\left[\mathcal{T}+\frac{1}{2\pi G}\left(\frac{T}{2}+\Lambda\right)\right]\label{LCDM1}\,.
\end{eqnarray} 
Integrating this differential equation we obtain the following action function
\begin{eqnarray}
f(T,\mathcal{T})=-2\Lambda+\sqrt{-T}F_g\left[\frac{T^{1/3}}{16\pi G}\left(16\pi G\mathcal{T}+T+8\Lambda\right)\right]\label{fLCDM}\,,
\end{eqnarray}
where $F_g[x]$ is a generic arbitrary function of its argument $x$. Taking into account the constraint (\ref{eqnonconserv2.1}), we see that $F_g[x]\equiv x$, linear function in $x$. The function $F_g[x]$ can also take a constant value. We see that the reconstruction of the action  (\ref{action2}) is given by  
\begin{eqnarray}
S&=&\frac{1}{16\pi G}\int d^4x e \left[T+f(T,\mathcal{T})+16\pi G\mathcal{L}_m\right]\nonumber\\
&=&\frac{1}{16\pi G}\int d^4x e \left\{T-2\Lambda+\sqrt{-T}\left[\frac{T^{1/3}}{16\pi G}\left(16\pi G\mathcal{T}+T+8\Lambda\right)\right]+16\pi G\mathcal{L}_m\right\}\label{action3}\,.
\end{eqnarray}
We can now see that this action generalizes the  $f(T)$ theory, because being able to have terms representing the interaction between the torsion and matter,  generated by the function $F_g$ in (\ref{fLCDM}). We also see clearly that this generalization is not for any functional form between $T$ and $\mathcal{T}$, but only by the product given in the third term of the above action. This restricts a lot the functional form of the theory, when it is imposes the validity of the $\Lambda$CDM model, as we have here. Another important observation is that this action lies in the $f(T)$ when we properly choose $F_g\left[(T^{1/3}/16\pi G)\left(16\pi G\mathcal{T}+T+8\Lambda\right)\right]\equiv Q$, where $Q$ is a constant given in \cite{salako}. In the next section, we will use  a model for this generic function, where $F_g[y]=y$, with $y=(T^{1/3}/16\pi G)\left(16\pi G\mathcal{T}+T+8\Lambda\right)$.
\par 
Here it is still some important observations. Our model is able to reproduce well the eras that the evolution of our universe goes. For exemple, with the imposition (\ref{LCDM}) we see daily that the era of radiation, where we write now $3H^2\approx 8\pi G\rho_r+\Lambda$, and of the matter, where $3H^2\approx 8\pi G\rho_m+\Lambda$, are now well reproduced. In radiation era, as the term of the radiation density dominates over the matter term, the generic function is being approximated by $F_g\left[(T^{1/3}/16\pi G)\left(T+8\Lambda\right)\right]$. Already in the matter era, as now the dominant term is the matter density, we have the same functional form of the action (\ref{action3}). In the dark energy era, it's approach $3H^2\approx \Lambda$, or equivalently $-(T/2)\approx \Lambda$, Which brings us to the same functional form of radiation era for $F_g$, but now $T\approx -2\Lambda$. Here inflation can be described similarly, whereas the component that dominates is the vacuum energy density.
\par 
{\bf As seen, our reconstruction is compatible with the most varied eras of cosmological evolution of our universe, leading us to believe that this model should also be compatible with the cosmological experimental data. We know at least that the acceleration of the universe at this stage is consistent with our model, for the very reconstruction is done to satisfy this condition. And we still have compatibility with the measures of type IA SN, where $\omega_{DE}\sim -1$ \cite{SIA}. We can at least see that there is a compatibility with ultra-stiff fluid where $\omega_m=1$, made in a recent analysis \cite{Diego2}.} 
\par 
One last important observation is that for the reconstruction satisfy the first constraint (\ref{eqnonconserv1.1}), we have to fix the trace in terms of torsion scalar as follows
\begin{eqnarray}
\mathcal{T}=\frac{1}{16\pi G}\left[\left(11+46¨08G^2\pi^2\right)T-8\Lambda\right]\label{Tcal}\,.
\end{eqnarray}
This shows us that the validity of the reconstruction of the $\Lambda$CDM model is subject to a fixation of the energy-momentum tensor trace, given in (\ref{Tcal}). This show us that the $f(T,\mathcal{T})$ theory should always be considered a theory in which the trace of energy-momentum tensor depends linearly on the torsion scalar. However, this does not permit us to conclude that the $f(T,\mathcal{T})$ Gravity must always fall back on $f(T)$ Gravity, because for what $f(T,\mathcal{T})$ fall back in $f(T)$, it is necessary that $F_g\equiv Q$ in (\ref{fLCDM}), and does not have a linear dependence on the trace, as we have here.
\par 
{\bf It may be possible to escape from the result that the action should have linear dependence on the trace $\mathcal{T}$, which is a direct implication of the conservation of the energy-momentum tensor given in \eqref{eqnonconserv2.1}. We can think about the possibility of non-conservation of energy-momentum tensor, arising from the non-minimal coupling between matter and torsion, similar to the case of theories modified only curvature \cite{Harko2,Harko3}. This should be interpreted as a possibility that the movement of the material content is not more geodesic having the appearance of an extra force term. This can also come to modify the bending of light and add a term in acceleration exerted by gravity, astrophysical implications may have to help in models of dark matter. This approach goes beyond our analysis here and should be done in future work.}
\par 
In the next section, we will show the stability of the  $\Lambda$CDM model reconstructed here.

\section{Stability of de Sitter and power-law solutions}\label{sec4}

In order to establish the validity of the  $\Lambda$CDM model, let us make a simple stability analysis. To do so, we will do a small perturbation in the geometry and the matter
\begin{eqnarray}
H(t)=h(t)[1+\delta (t)]\;,\;\rho_{mt}(t)=\rho_{mth}(t)[1+\delta_m(t)]\label{perturbation}\,,
\end{eqnarray}
where $\{h(t),\rho_{mth}(t)\}$ is an exact solution of the equations of motion for call background  (obey to  (\ref{eqm3})-(\ref{eqm4})) and $\delta(t),\delta_m(t)<<1$. With this, the torsion scalar is given by 
\begin{eqnarray}
T=-6H^2=-6h^2(1+\delta)^2=T_0(t)[1+2\delta(t)]\,,\label{st1}
\end{eqnarray}
where we have  $T_0(t)=-6h^2(t)$. The trace of the energy-momentum tensor is given by 
\begin{eqnarray}
\mathcal{T}(t)=\rho_{mt}(t)-3p_{mt}(t)=(1-3\omega_m)\rho_{m}(t)+(1-3\omega_r)\rho_{r}(t)=\rho_{m}(t)=\mathcal{T}_0(t)[1+\delta_m(t)]\,,
\end{eqnarray} 
where $\mathcal{T}_0(t)=\rho_{mh}(t)$. The perturbations of the torsion scalar and the trace of the energy-momentum tensor are 
\begin{eqnarray}
\bar{\delta}T(t)=T-T_0=2T_0(t)\delta(t)\;,\;\bar{\delta}\mathcal{T}(t)=\mathcal{T}-\mathcal{T}_0=\mathcal{T}_0(t)\delta_m(t)\,\label{perturbation2}\,.
\end{eqnarray}

Let us express the function  $f(T,\mathcal{T})$ in the  Taylor series until the first order around the point  $[T_0(t),\mathcal{T}_0(t)]$, by 
\begin{eqnarray}
&&f(T,\mathcal{T})\approx f(T_0,\mathcal{T}_0)+\left[\frac{\partial f}{\partial T}\right]_{T=T_0}\bar{\delta}T+\left[\frac{\partial f}{\partial \mathcal{T}}\right]_{\mathcal{T}=\mathcal{T}_0}\bar{\delta}\mathcal{T}\,.\label{pf}
\end{eqnarray}
its derivatives, that are obtained deriving (\ref{pf}), can be expressed as 
\begin{eqnarray}
&&f_T(T,\mathcal{T})\approx f_T(T_0,\mathcal{T}_0)+\left[\frac{\partial^2 f}{\partial T^2}\right]_{T=T_0}\bar{\delta}T+\left[\frac{\partial^2 f}{\partial T\partial \mathcal{T}}\right]_{T=T_0,\mathcal{T}=\mathcal{T}_0}\bar{\delta}\mathcal{T}\,,\\
&&f_{\mathcal{T}}(T,\mathcal{T})\approx f_{\mathcal{T}}(T_0,\mathcal{T}_0)+\left[\frac{\partial^2 f}{\partial T \partial \mathcal{T}}\right]_{T=T_0,\mathcal{T}=\mathcal{T}_0}\bar{\delta}T+\left[\frac{\partial^2 f}{\partial \mathcal{T}^2}\right]_{\mathcal{T}=\mathcal{T}_0}\bar{\delta}\mathcal{T}\,.
\end{eqnarray}
By substituting these expressions into (\ref{eqm3}), collecting the term until the first order, we get the following result 
\begin{eqnarray}
&&3h^2(1+2\delta)=8\pi G\rho_{mth}(1+\delta_m)-\frac{1}{2}\Bigg\{\left[f_0+f_{T_0}\bar{\delta}T+f_{\mathcal{T}_0}\bar{\delta}\mathcal{T}\right]+12h^2(1+2\delta)\times\nonumber\\
&&\times\left[f_{T_0}+f_{T_0T_0}\bar{\delta}T+f_{T_0\mathcal{T}_0}\bar{\delta}\mathcal{T}\right]+\Bigg\}+\left[f_{\mathcal{T}_0}+f_{T_0\mathcal{T}_0}\bar{\delta}T+f_{\mathcal{T}_0\mathcal{T}_0}\bar{\delta}\mathcal{T}\right]\left(\rho_{mh}+\frac{4}{3}\rho_{rh}\right)\left(1+\delta_m\right)\nonumber\\
\end{eqnarray}
that replacing $\rho_{mh}=\mathcal{T}_0$, $h_0^2=-T_0/6$ and $\rho_{rh}=-\left[(1/8\pi G)((T_0/2)+\Lambda)+\mathcal{T}_0\right]$ becomes
\begin{eqnarray}
&&-T_0\left\{1+f_{T_0}+2T_0f_{T_0T_0}-\frac{2}{3}\left[\mathcal{T}_0+\frac{1}{2\pi G}\left(\frac{T_0}{2}+\Lambda\right)\right]f_{T_0\mathcal{T}_0}\right\}\delta(t)=\Bigg\{\frac{T_0}{2}+\Lambda\nonumber\\
&&-\frac{1}{6}\left[5\mathcal{T}_0+\frac{1}{\pi G}\left(\frac{T_0}{2}+\Lambda\right)\right]f_{\mathcal{T}_0}+T_0\mathcal{T}_0f_{T_0\mathcal{T}_0}-\frac{1}{3}\mathcal{T}_0\left[\mathcal{T}_0+\frac{1}{2\pi G}\left(\frac{T_0}{2}+\Lambda\right)\right]f_{\mathcal{T}_0\mathcal{T}_0}\Bigg\}\delta_m(t)\,.\label{eq2}
\end{eqnarray}
We maintain for the moment this result. Now we have to express the perturbation  $\delta(t)$  in terms of  $\delta_m(t)$. To do so, let us take the perturbation of the conservation equation of the energy-momentum tensor 
\begin{eqnarray}
&&\dot{\rho}_{mt}+3H\left(\rho_{mt}+p_{mt}\right)=0\\
&&\dot{\rho}_{mth}(1+\delta_m)+\rho_{mth}\dot{\delta}_m+3h(1+\delta)\left(\rho_{mh}+\frac{4}{3}\rho_{rh}\right)(1+\delta_m)=0\nonumber\\
&&\delta(t)=-\frac{\rho_{mth}}{3h\left(\rho_{mth}+p_{mth}\right)}\dot{\delta}_m(t)=-\frac{\sqrt{6}}{8\pi G\sqrt{-T_0}}\frac{\left[(T_0/2)+\Lambda\right]}{\left\{(1/8\pi G)[(T_0/2)+\Lambda]+\mathcal{T}_0\right\}}\dot{\delta}_m(t)\label{pconserv}\,.
\end{eqnarray}
Here, the ``dot" denotes the derivative with respect to time $t$. Now, by inserting  (\ref{pconserv}) in (\ref{eq2}), one gets 
\begin{eqnarray}
&&\delta_m=\delta_m^{(0)}\exp\left[\int\frac{F_1(t)}{F_2(t)}dt\right]\label{delm}\,,\\
&&F_1(t)=\frac{T_0}{2}+\Lambda-\frac{1}{6}\left[5\mathcal{T}_0+\frac{1}{\pi G}\left(\frac{T_0}{2}+\Lambda\right)\right]f_{\mathcal{T}_0}+T_0\mathcal{T}_0f_{T_0\mathcal{T}_0}-\frac{1}{3}\mathcal{T}_0\left[\mathcal{T}_0+\frac{1}{2\pi G}\left(\frac{T_0}{2}+\Lambda\right)\right]f_{\mathcal{T}_0\mathcal{T}_0}\,,\\
&&F_2(t)=\sqrt{-T_0}\left(\sqrt{6}T_0+2\Lambda\right)\left\{-6\pi G \left[1+f_{T_0}+2T_0f_{T_0T_0}\right]+\left(4\pi G\mathcal{T}_0+T_0+2\Lambda\right)f_{T_0\mathcal{T}_0}\right\}\Big/\Big[24\pi G\times\nonumber\\
&&\times\left(4\pi G\mathcal{T}_0+T_0+2\Lambda\right)\Big]\,.
\end{eqnarray}

This equation is valid for specific  models, the ones we have chosen, the de Sitter and  power-law models. We will substitute the characteristic of these models in the above equation. 
\par 
Let us first analyse the stability of de Sitter solution, where the characteristic is given by  $h=h_0$,  yielding through the equation of continuity
\begin{eqnarray}
\rho_{mh}(t)=\rho_0\exp[-3h_0t]\label{dens1}\,.
\end{eqnarray}
The torsion scalar is constant and is given by $T_0=-6h_0^2$. The trace of the energy-momentum tensor is given by 
\begin{eqnarray}
\mathcal{T}_0=\rho_0\exp[-3h_0t]\label{trace1}\,.
\end{eqnarray}

Now we have to use the reconstruction of $f(T,\mathcal{T})$ for $\Lambda$CDM in (\ref{fLCDM}), specifying the case of de Sitter solution. We first consider the time depending torsion scalar, i.e, variable torsion scalar, and then fix for the de Sitter solution, we can then correctly calculate the derivatives of the function  $f(T, \mathcal{T})$.

For the de Sitter solution, we take one case for action (\ref{action3}). We make the  case where $F_g[y]=y$, with $y=(T_0^{1/3}/16\pi G)\left(16\pi G\mathcal{T}_0+T_0+8\Lambda\right)$.
\par 
For this case, we have the following integral in (\ref{delm})
\begin{eqnarray}
&&\delta_m(t)=\delta_m^{(0)}\exp\Bigg\{-\frac{4\times 6^{1/3}(h_0^{5/3}+6^{1/6}\pi G)(3h_0^2-\Lambda)\left[(3h_0^2-\Lambda)t+\frac{2\pi G\rho_0}{3h_0}e^{-3h_0 t} \right]}{h_0^{2/3}(6^{5/6}h_0^2-6\pi G h_0^{1/3})(3\sqrt{6}h_0^2-\Lambda)}\Bigg\}\label{delmd}
\end{eqnarray}
and from (\ref{pconserv})
\begin{eqnarray}
\delta (t)=-\delta_m(t)\frac{4\times 6^{1/3}}{h_0^{5/3}}\frac{(h_0^{5/3}+6^{1/6}\pi G)(3h_0^2-\Lambda)\left[3h_0^2-\Lambda-2\pi G\rho_0 e^{-3h_0 t}\right]}{(6^{5/6}h_0^2-6\pi G h_0^{1/3})(3\sqrt{6}h_0^2-\Lambda)}\,.\label{deld}
\end{eqnarray}
 
\begin{figure}[h]
\centering
\begin{tabular}{rl}
\includegraphics[height=4cm,width=8cm]{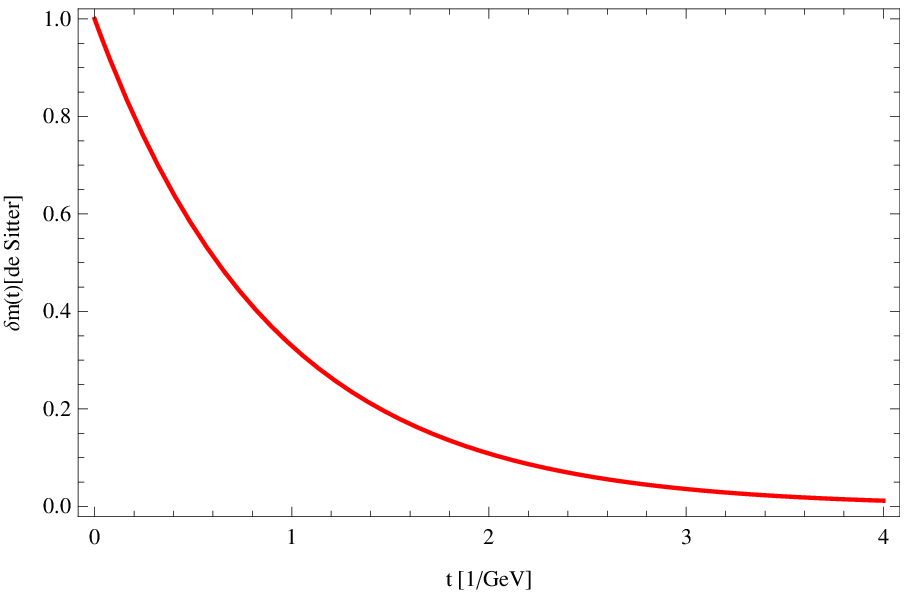}&
\includegraphics[height=4cm,width=8cm]{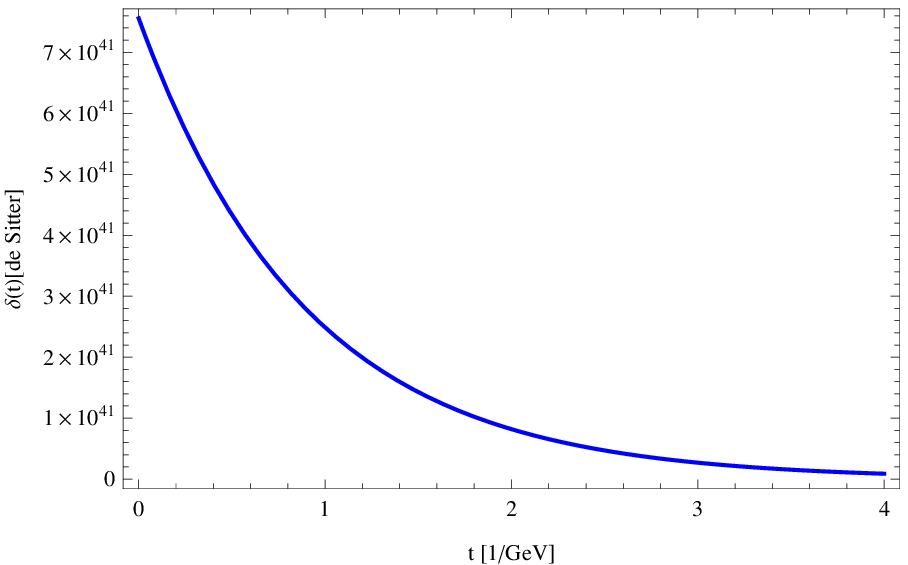}
\end{tabular}
\caption{\scriptsize{The left graph shows the temporal evolution of $\delta_{m}(t)$ (in red) and the of the right the temporal evolution of $\delta_m(t)$ (in blue) both for the case $F_g[y]=y$, with $y=(T_0^{1/3}/16\pi G)\left(16\pi G\mathcal{T}_0+T_0+8\Lambda\right)$, the two for the particular case of de Sitter solution. The parameters are chosen as 
$\{N=20, h_0=2.1\times 0.7\times 10^{-42}, \delta^{(0)}_m=1,\rho_0=0.1\times 10^{-121}, G=(1.2)^2\times 10^{-38},\Lambda=10^{-42}\}$. }}
\label{fig1}
\end{figure}

We do a numerical graph representing the temporal evolution of these perturbations in Figure \ref{fig1}. We can see that the perturbations rapidly decrease to zero, showing a possible stability for the solution of de Sitter. There is a serious problem here. The graphical representation to $\delta(t)$ in figure \ref{fig1} shows that despite this function quickly drop to zero, the initial values are of the order to $10^{41}$, thus contradicting the initial assumption for perturbation which is $\delta(t)<<1$. This shows us an impossibility for stability to the model of de Sitter, due precisely to the initial values for perturbation of the geometric part.
\par
Doing the same procedure for the case of power law, where now $h_0(t)=\alpha/t$, the equation (\ref{delm}), for the case $F_g=y$ with $y=(T_0^{1/3}/16\pi G)\left(16\pi G\mathcal{T}_0+T_0+8\Lambda\right)$, becomes 
\begin{eqnarray}
&&\delta_m(t)=\delta_m^{(0)}\exp\Bigg\{\int \frac{F_{pl1}}{F_{pl2}}dt\Bigg\}\label{delmp1}\\
&&F_{pl1}=-4\times 6^{1/3}\left(\frac{t}{t_0}\right)^{-3\alpha}\frac{\alpha}{t}\left[\alpha^2+6^{1/6}\pi G t^2\left(\frac{\alpha}{t}\right)^{1/3}\right](\Lambda t^2-3\alpha^2)\left[(\Lambda t^2-3\alpha^2)\left(\frac{t}{t_0}\right)^{3\alpha}+2\pi G\rho_0 t^2\right]\,,\\
&&F_{pl2}=\alpha^2(3\sqrt{6}\alpha^2-\Lambda t^2)\left[6^{5/6}\alpha^2-6\pi G t^2\left(\frac{\alpha}{t}\right)^{1/3}\right].
\end{eqnarray}

\begin{figure}[h]
\centering
\begin{tabular}{rl}
\includegraphics[height=4cm,width=8cm]{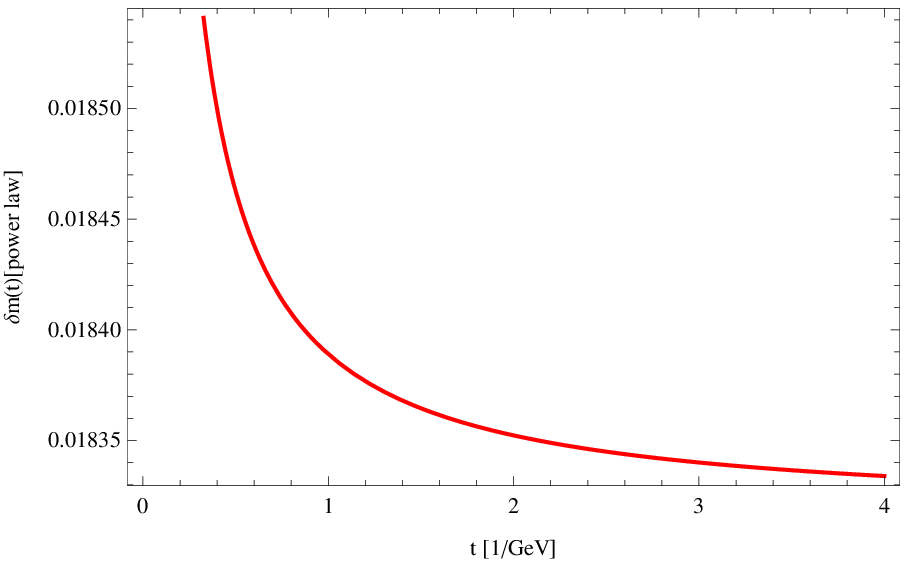}&
\includegraphics[height=4cm,width=8cm]{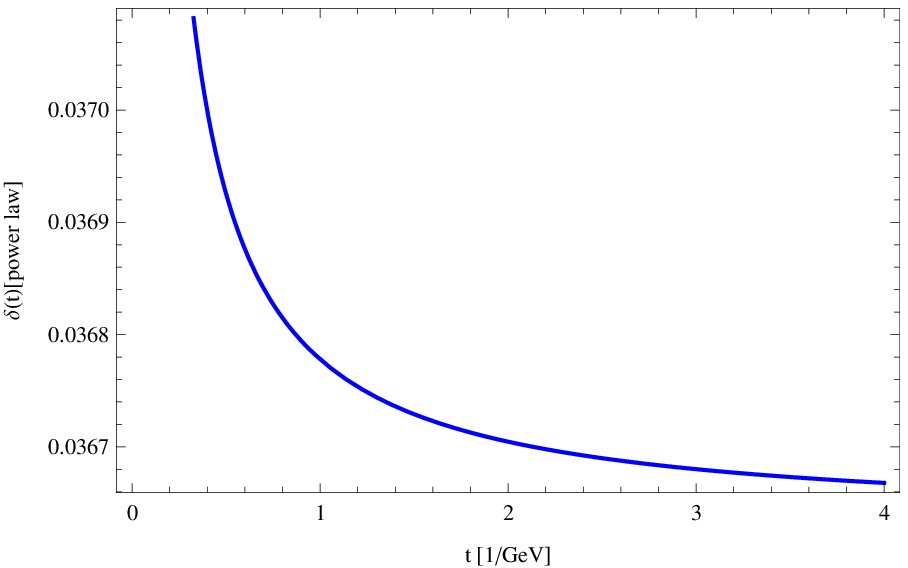}
\end{tabular}
\caption{\scriptsize{The left graph shows the temporal evolution of $\delta_m(t)$ (in red) and the of the right for temporal evolution of $\delta(t)$ (in blue) both for the case $F_g[y]=y$, with $y=(T_0^{1/3}/16\pi G)\left(16\pi G\mathcal{T}_0+T_0+8\Lambda\right)$,  the two for the particular case of power law solution. The parameters are chosen as 
$\{\alpha=2, t_0=3, h_0=2.1\times 0.7\times 10^{-42}, \delta^{(0)}_m=1,\rho_0=0.1\times 10^{-121}, G=(1.2)^2\times 10^{-38},\Lambda=10^{-42}\}$. }}
\label{fig2}
\end{figure}

From (\ref{pconserv}), with (\ref{delmp1}), we find
\begin{eqnarray}
&&\delta(t)=\delta_m(t)\frac{F_{pl1}}{F_{pl2}}\left[\frac{\Lambda t^2-3\alpha^2}{\alpha t \left(\Lambda-\frac{3\alpha^2}{t^2}+8\pi G\rho_0\left(\frac{t}{t_0}\right)^{-3\alpha}\right)}\right]
\end{eqnarray}
We numerically represent the evolution of perturbations $\delta_m(t)$ and $\delta(t)$ in the figure \ref{fig2}. Again we can see that the perturbations quickly decay to zero, showing the stability of this solution.
\par 
In the next section, we will study the two laws of thermodynamics for the $\Lambda$CDM model.


\section{Thermodynamic Laws according to $\Lambda$CDM model}\label{sec5}
By analogy to  the thermodynamics of black holes, the thermodynamics of the cosmological models can be realised, verifying if these models satisfy the classical thermodynamics. This has been done in the context of GR \cite{termGR}. It has also been of wide interest the study of these laws for the case of the theories of modified gravities \cite{termmod}. In the context of modified gravity it is common to use a description of non-equilibrium thermodynamics for representing the physical system  \cite{non-eq}, we will use this representation here.
\par   
In this section we will investigate the conditions of satisfying the first and second laws of thermodynamics.
\subsection{First law}
We will redefine now the action of the theory $f(T,\mathcal{T})$ as  
\begin{eqnarray}
S=\int d^4x e \left[f(T,\mathcal{T})+16\pi G\mathcal{L}_m\right]\label{newaction}.
\end{eqnarray}
This is necessary to get the possibility for defining an effective Newton constant $G_{eff}$. Now the equations of motion are rewritten as 
\begin{eqnarray}
&&3H^2=8\pi G_{eff}\left(\rho_{mt}+\rho_{DE}\right)\,,\label{DE1}\\
&&\dot{H}=-4\pi G_{eff}\left(\rho_{mt}+p_{mt}+\rho_{DE}+p_{DE}\right)\label{DE2}\,,
\end{eqnarray}
with 
\begin{eqnarray}
&&G_{eff}=\frac{G}{f_T}\left(1+\frac{f_{\mathcal{T}}}{16\pi G}\right)\label{Geff}\,,\\
&&\rho_{DE}=\frac{1}{16\pi G_{eff}f_T}\left(f_{\mathcal{T}}p_{mt}-\frac{1}{2}f\right)\label{densDE}\,,\\
&&p_{DE}=\rho_{mt}+p_{mt}-\rho_{DE}-\frac{1}{4\pi G_{eff}f_T}\left[12H^2\dot{H}f_{TT}-f_{T\mathcal{T}}H\left(\dot{\rho}_{mt}-3\dot{p}_{mt}\right)\right]\label{pDE}\,.
\end{eqnarray}

We impose the conservation of the matter sector  $\dot{\rho}_{mt}+3H(\rho_{mt}+p_{mt})=0$, as discussed in the sections \ref{sec2.1} and \ref{sec3}. But as we are doing a non-equilibrium thermodynamics description, the dark energy sector is does not conserve,
\begin{eqnarray}
\dot{\rho}_{DE}+3H(\rho_{DE}+p_{DE})=\frac{3H^2}{8\pi}\frac{d}{dt}\left(\frac{1}{G_{eff}}\right)\label{conservDE}\,,
\end{eqnarray}
where we used  (\ref{DE1})-(\ref{pDE}). This concords with the theory  analogue to  $f(T,\mathcal{T})$, in \cite{sharif2},  and also for the particular case where  $f\equiv f(T)$ in \cite{bamba2}. We still should compare our results with the ones of  $f(R)$ theory, in \cite{bamba3}. It is also clear here that, when $G_{eff}=G$ (in the particular case  $f(T,\mathcal{T})=T$) the dark energy sector is conserved, the TT is recovered, or the GR analogously, within an equilibrium description of thermodynamics.
\par 
Now, we establish the basic tools for the first law of thermodynamics. The line element (\ref{FLRW}) can be rewritten as 
\begin{eqnarray}
dS^2=h_{\mu\nu}dx^{\mu}dx^{\nu}+\widehat{r}^2\left(d\theta^2+\sin^2\theta d\phi^2\right)\,,\label{newFLRW}
\end{eqnarray}
where we have the metric of 2-dimensional space $[h_{\mu\nu}]=[1,-a^2(t)]$, for  $\mu,\nu=0,1$ and $x^1=t,x^2=r$, a new radial coordinate $\widehat{r}=ra(t)$, with $r$ being the usual radial coordinate $r^2=x^2+y^2+z^2$. Here the space-time can be decomposed in a 2-dimensional space with the metric  $h_{\mu\nu}$ and other  2-dimensional space with is a 2-sphere, with the line element $d\Omega^2=d\theta^2+\sin^2\theta d\phi^2$.
\par 
Through this line element  (\ref{newFLRW}), we can calculate the apparent horizon\footnote{We choose to  work the apparent horizon due to the suggestion according to what the event horizon  does not satisfy to the second law of thermodynamics, contrary to the apparent one where the law is satisfied \cite{zhou}.} and the associated temperature to this. The apparent horizon is obtained from the expression $h^{\mu\nu}\partial_{\mu}\widehat{r}_A\partial_{\nu}\widehat{r}_A=0$, which results in 
\begin{eqnarray}
\widehat{r}_A=\frac{1}{H}\label{AH}.
\end{eqnarray}
The temperature is calculated by the expression 
\begin{eqnarray}
T_A=\frac{1}{4\pi \sqrt{-h}}\left|\partial_{\mu}\left[\sqrt{-h}h^{\mu\nu}\partial_{\nu}\widehat{r}_A\right]\right|
\end{eqnarray}
where $h=det[h_{\mu\nu}]$. The temperature is given by 
\begin{eqnarray}
T_A=\frac{\widehat{r}_A}{4\pi}\left(\dot{H}+2H^2\right)\label{temp},
\end{eqnarray}
where, from now, we impose $\dot{H}+2H^2>0$ for getting a definite positive temperature.
\par
Using (\ref{AH}) and  (\ref{DE2}), we can calculate the derivative of $\widehat{r}_A$
\begin{eqnarray}
\frac{d\widehat{r}_A}{dt}=4\pi G_{eff}\widehat{r}_A^2\left(\rho_{tot}+p_{tot}\right)\,,\label{dAH}
\end{eqnarray}
with $\rho_{tot}=\rho_{mt}+\rho_{DE}$ and  $p_{tot}=p_{mt}+p_{DE}$. Now, we will take the entropy related to apparent horizon, for modified gravity theories \cite{brustein}, that is $S_A=A/(4G_{eff})$, with $A=4\pi \widehat{r}_A^2$. Here, appears a new restriction for the thermodynamic system, $G_{eff}>0$ for getting a definite positive entropy $S_A\geq0$. Considering this entropy, and using $G_{eff}$ in (\ref{Geff}) and  (\ref{dAH}),  we get the following differential
\begin{eqnarray}
dS_A=2\pi \widehat{r}_A\left[4\pi \widehat{r}_A^2\left(\rho_{tot}+p_{tot}\right)dt+\frac{1}{2}\widehat{r}_Ad\left(\frac{1}{G_{eff}}\right)\right]\label{dSA}\,.
\end{eqnarray}
Now we take the  Misner-Sharp energy for modified gravities $E_{MS}=\widehat{r}_A/(2G_{eff})=V\rho_{tot}$\footnote{Once again we have to restrict $G_{eff}>0$ for obtaining definite positive energy. This has been indicated in \cite{starobinsky}, in  order to avoid ghost structure in the quantised theory.} \cite{cai1}, with $V=(4/3)\pi \widehat{r}_A^3$, where, using  (\ref{dAH}), the differential gives rise to
\begin{eqnarray}
&&dE_{MS}=2\pi \widehat{r}_A^2\left(\rho_{tot}+p_{tot}\right)dt+\frac{1}{2}\widehat{r}_A d\left(\frac{1}{G_{eff}}\right)\label{MSe}\,.
\end{eqnarray}

Using (\ref{AH}) in (\ref{temp}), considering  (\ref{dAH}), we can rewrite the temperature as 
\begin{eqnarray}
T_A=\frac{\widehat{r}_A}{4\pi}\left[\frac{d}{dt}\left(\frac{1}{\widehat{r}_A}\right)+2\left(\frac{1}{\widehat{r}_A}\right)^2\right]=\frac{1}{2\pi\widehat{r}_A}\left(1-\frac{1}{2}\frac{d\widehat{r}_A}{dt}\right)\label{temp2}\,.
\end{eqnarray}
Making the product (\ref{temp2}) for (\ref{dSA}) we have 
\begin{eqnarray}
T_AdS_A&=&-dE_{MS}+2\pi\widehat{r}_A^2\left(\rho_{tot}-p_{tot}\right)d\widehat{r}_A+6\pi\widehat{r}_A^2\left(\rho_{tot}+p_{tot}\right)dt-4\pi\widehat{r}_A^2\rho_td\widehat{r}_A\nonumber\\
&+&\frac{1}{2}\widehat{r}_A\left(1+2\pi\widehat{r}_AT_A\right)d\left(\frac{1}{G_{eff}}\right)\label{TdSA}\,.
\end{eqnarray}
Defining the work density given as  $W=(1/2)(\rho_{tot}-p_{tot})$ \cite{bamba2}, using (\ref{DE1}) and  (\ref{dAH}), we can rewrite (\ref{TdSA}) as
\begin{eqnarray}
&&T_A dS_A+T_AdS_p=-dE_{MS}+WdV\label{firstlaw}\,,\\
&&T_AdS_p=-\frac{1}{2}\widehat{r}_A\left(1+2\pi\widehat{r}_AT_A\right)d\left(\frac{1}{G_{eff}}\right)\label{Sp}\,,
\end{eqnarray}
where the term $T_AdS_p$ can be interpreted as coming from an entropy production. Considering $T_A>0$, we eliminate the singularity in the entropy production in  $\dot{H}=-2H^2$ ($T_A=0$). Here, we see that the result concords with the particular case where $f\equiv f(T)$, in \cite{bamba2}. Also in the analogue theory $f(R,\mathcal{T})$ in  \cite{sharif2}. This entropy production $dS_p$ is directly linked to the non-conservation of the dark sector in  (\ref{conservDE}). When we fix the particular case $f(T,\mathcal{T})=T$ ($G_{eff}=G$), we recover a theory without entropy production, i.e, the TT.
\par
\begin{figure}[h]
\centering
\begin{tabular}{rl}
\includegraphics[height=4cm,width=9cm]{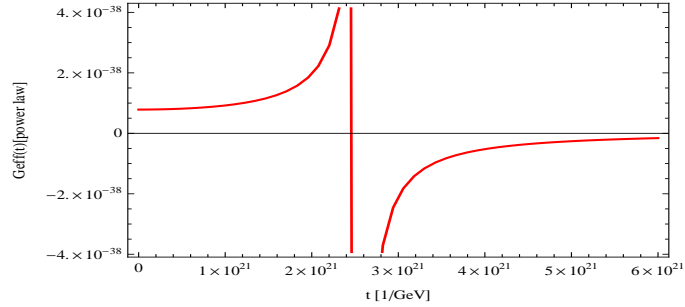}
\end{tabular}
\caption{\scriptsize{The graph shows the temporal evolution of $G_{eff}(t)$ for the case $F_g[y]=y$, with $y=(T_0^{1/3}/16\pi G)\left(16\pi G\mathcal{T}_0+T_0+8\Lambda\right)$, for the particular case of power law solution. The parameters are chosen as 
$\{\alpha=2, t_0=3, h_0=2.1\times 0.7\times 10^{-42}, \delta^{(0)}_m=1,\rho_0=0.1\times 10^{-121}, G=(1.2)^2\times 10^{-38},\Lambda=10^{-42}\}$. }}
\label{fig3}
\end{figure}

 
We take the effective constant (\ref{Geff}) to the power law solution, which provides the
\begin{eqnarray}
G_{eff}(t)=\frac{6G\left(\frac{t}{t_0}\right)^{3\alpha}\left[8\times 6^{1/6}\pi Gt^2\left(\frac{\alpha}{t}\right)^{1/3}-3\alpha^2\right]}{\left[20t^2\left(\Lambda\left(\frac{t}{t_0}\right)^{3\alpha}+2\pi G\rho_0\right)-33\alpha^2\left(\frac{t}{t_0}\right)^{3\alpha}\right]}
\end{eqnarray}
We represent the temporal evolution of the effective Newton constant $G_{eff}(t)$ in the figure \ref{fig3}, we show what for the case of power law, $G_{eff}$ assumes negative values, precluding a definition of  Misner-Sharp energy and entropy positives.

So we can conclude that this type of behavior suggests (it we have not tested all possibilities of cases) that the $f(T,\mathcal{T})$ Gravity, does not have a well-established thermodynamic with all the thermodynamic quantities physical well-defined every time interval.

\subsection{Second law}\label{2Law}

We will impose the second law of thermodynamics to  $f(T,\mathcal{T})$. To do so, let us consider the Gibbs equation in terms of matter fluid and total energy given by 
\begin{eqnarray}
T_{tot}dS_{tot}=\left(\rho_{tot}+p_{tot}\right)dV+Vd\rho_{tot}\,,\label{Gibbs}
\end{eqnarray}
where $T_{tot}$ and  $S_{tot}$ are the temperature and total entropy  associated to the system, respectively. According to the second law of thermodynamics, we must have 
\begin{eqnarray}
\frac{dS_{tot}}{dt}+\frac{dS_A}{dt}+\frac{dS_p}{dt} \geq 0\,. \label{2LT}
\end{eqnarray}
Knowing that $E_{MS}=V\rho_{tot}$ and using (\ref{firstlaw}) and  (\ref{Gibbs}) we can write the left side of  (\ref{2LT}) as 
\begin{eqnarray}
\frac{dS_{tot}}{dt}+\frac{dS_A}{dt}+\frac{dS_p}{dt}=\frac{1}{T_AT_{tot}}\left[\left(\rho_{tot}+p_{tot}\right)\frac{dV}{dt}\left(T_A-\frac{T_{tot}}{2}\right)+V\frac{d\rho_{tot}}{dt}\left(T_A-T_{tot}\right)\right]\,.\label{2LT2}
\end{eqnarray}
For $T_A=T_{tot}$, using (\ref{DE2}) and  $dV=4\pi\widehat{r}^2_Ad\widehat{r}_A$ we get, for (\ref{2LT2}), 
\begin{eqnarray}
\frac{dS_{tot}}{dt}+\frac{dS_A}{dt}+\frac{dS_p}{dt}=\frac{1}{2T_AG_{eff}}\left(\frac{d\widehat{r}_A}{dt}\right)^2\,,\label{2LT3}
\end{eqnarray}
with $G_{eff}> 0$, as previously done. The observation is that the temperature of the  apparent horizon in  (\ref{temp}), depends on the components of the universe, through  (\ref{DE1}) and  (\ref{DE2}), including the photons of CMB with $2.73K$. Assuming that the total temperature of the system $T_{tot}$ is equal to the one of the apparent horizon, i.e, this temperature also includes the photons of CMB. 
\par 
Here are the expressions of the second law similar to particular cases, or analogs $f(T)$ and $f(R,\mathcal{T})$, but we are still restricted to the problem of the previous section, in which we can not set $G_{eff}$ always positive, which leads to a possibility of negative entropy, which is physically inconsistent for the cases studied here.

\section{Comparison with previous results}\label{sec5.1}
We noticed a great need to compare our results with the first work in this extension of the $f(T)$ Gravity. The observation that we have done here is that the first construction of $f(T,\mathcal{T})$ theory has been done by  Kiani and Nozari \cite{kiani}. But do to the ambiguity of the choice of the Lagrangian density for the perfect fluid, with the Lagrangian density being  $\mathcal{L}_m=p$ (our choice) or  $\mathcal{L}_m=-\rho$ (see \cite{brown}), they use the last description. This conduces to a new approach possibly equivalent to the same theory. They make an analysis of the stability for the de Sitter model, but as they did not reconstruct the action, they considered the model as a generic one $f(T,\mathcal{T})=k_1T+k_2T^m\mathcal{T}^n$. 
\par 
Here we will verify the consistency of this approach to the conservation of energy-momentum tensor. Taking the equation of motion (11) in \cite{kiani}, considering (13), we multiply by $e^a_{\;\;\nu}$ and considering the identity $e^{a}_{\;\;\nu}[e^{-1}\partial_{\lambda}(ee_{a}^{\;\;\alpha}S_{\alpha}^{\;\;\sigma\lambda})-e_{a}^{\;\;\alpha}T^{\lambda}_{\;\;\gamma\alpha}S_{\lambda}^{\;\;\gamma\sigma}]=(1/2)[G_{\nu}^{\;\;\mu}-(1/2)\delta^{\mu}_{\nu}T]$, we have 
\begin{eqnarray}
S_{\nu}^{\;\;\mu\sigma}f_{TT}\partial_{\sigma}T+\frac{1}{2}f_T\left(G_{\nu}^{\;\;\mu}-\frac{1}{2}\delta^{\mu}_{\nu}T\right)+\delta^{\mu}_{\nu}\left(\frac{1}{4}f+\frac{3}{2}\rho_{mt}f_{\mathcal{T}}\right)=\left(4\pi G+\frac{5}{4}f_{\mathcal{T}}\right)\Theta^{\;\;\mu}_{\nu}\label{eqkiani}\,.
\end{eqnarray}
Now we take the total divergence $\nabla_{\mu}$ of the above equation of motion, and isolating $\nabla_{\mu}\Theta_{\nu}^{\;\;\mu}$, taking into account that $\nabla_{\mu}G_{\nu}^{\;\;\mu}\equiv 0$, we have the following result
\begin{eqnarray}
\nabla_{\mu}\Theta_{\nu}^{\;\;\mu}&=&\frac{1}{\left(4\pi G+(5/4)f_{\mathcal{T}}\right)}\Bigg\{\nabla_{\mu}S_{\nu}^{\;\;\mu\sigma}f_{TT}\partial_{\sigma}T+S_{\nu}^{\;\;\mu\sigma}\left[f_{TT}\nabla_{\mu}\partial_{\sigma}T+\partial_{\sigma}T\left(f_{TTT}\partial_{\mu}T+f_{TT\mathcal{T}}\partial_{\mu}\mathcal{T}\right)\right]\nonumber\\
&&+\frac{1}{2}\left(f_{TT}\partial_{\mu}T+f_{T\mathcal{T}}\partial_{\mu}\mathcal{T}\right)\left(G_{\nu}^{\;\;\mu}-\frac{1}{2}\delta^{\mu}_{\nu}T\right)+\frac{1}{4}f_{\mathcal{T}}\partial_{\nu}\mathcal{T}+\frac{3}{2}\Big[f_{\mathcal{T}}\partial_{\nu}\rho_{mt}+\rho_{mt}\left(f_{T\mathcal{T}}\partial_{\nu}T+f_{\mathcal{T}\mathcal{T}}\partial_{\nu}\mathcal{T}\right)\Big]\nonumber\\
&&-\frac{5}{4}\left(f_{T\mathcal{T}}\partial_{\mu}T+f_{\mathcal{T}\mathcal{T}}\partial_{\mu}\mathcal{T}\right)\Theta_{\nu}^{\;\;\mu}\Bigg\}\,.
\end{eqnarray}
Then arise two constraint here, one for $\nu=0$ and outher for $\nu=1,2,3$, 
\begin{eqnarray}
\dot{\rho}_{mt}+3H(\rho_{mt}+p_{mt})&=&\frac{1}{4}\left\{\left[\rho_{mt}f_{T\mathcal{T}}+12(1-2H^2)f_{TT}\right]\frac{dT}{dt}+\left[\rho_{mt}f_{\mathcal{T}\mathcal{T}}+f_{\mathcal{T}}+12(1-H^2)f_{T\mathcal{T}}\right]\frac{d\mathcal{T}}{dt}+6f_{\mathcal{T}}\dot{\rho}_{mt}\right\}\label{nonconservkiani1}\,,\\
0&=&3(H^2-a)\left[f_{TT}\frac{dT}{dt}+f_{T\mathcal{T}}\frac{d\mathcal{T}}{dt}\right]\label{nonconservkiani2}\,.
\end{eqnarray}
The second constraint is satisfied, in general, to $(d/dt)f_T\equiv 0$, fixing function $f(T,\mathcal{T})$ as a linear function on the torsion scalar $T$. This returns to a trace coupling with Teleparallel Theory, which is not the intention here. Thus, the description of the theory as shown in \cite{kiani}, is not consistent with the conservation of energy-momentum tensor presented here. We must then discard here that description by the density of the fluid.
\par 
An important observation is that this result does not mean that the description of Lagrangian fluid through the density is incorrect, but that the specific choice of the set of diagonal tetrad, $[e^{b}_{\;\;\mu}]=[1,a(t),a(t),a(t)]$, is that it can not provide a consistent result. It should be investigated in another future work, as well as consistency between two description possibilities, one by density and  the other by fluid pressure.
 
\section{Conclusion}\label{sec6}

The $ f(T, \mathcal{T})$ theory can be a good opportunity to test whether we can  describe the dark energy by changing the TT equations of motion. To do this, it is first necessary to test the most common models of modern cosmology, the famous  $\Lambda$CDM one for exemple.
\par
We reconstruct the gravitational action of this model and have done a brief analysis of their stability. The result is resumed as follows.
\par
The reconstruction of $ \Lambda$CDM model gives rise to an action as combination of a linear term in the torsion scalar, a constant ($ -2 \Lambda$) and another, which is the interaction between matter and torsion, described by the product $\sqrt{-T}F_g\left[(T^{1/3}/16\pi G)\left(16\pi G\mathcal{T}+T+8\Lambda\right)\right]$. This reconstruction decays in the  $ f(T)$ theory, by appropriately choosing the $ F_g \equiv Q $, with $Q$ a constant given in \cite{salako}. This reconstruction severely restricted the functional form of action, but well reproduces the cosmological eras. 
\par 
{\bf We do the analysis of the energy-momentum tensor conservation through the obtained reconstruction model, resulting in the function $F_g[x]$ must be linear in its argument, so that conservation is satisfied as the usual $\Lambda$CDM model. With this action should not contain terms of higher orders and is restricted to linearity in the trace $\mathcal{T}$. This model does not prevent accordance with observational data, and for the particular case of  ultra-stiff fluid, where $\omega_m=1$ (see \cite{Diego2}).} 
\par
We have done the analysis of the stability of this model, for a particular choice $F_g=y$, and we found that the model is stable for de Sitter and power law solutions, except for geometric part to de Sitter one.
\par 
We finish analysing the thermodynamics for the  $\Lambda$CDM model. We find that the imposition of the thermodynamic laws are satisfied for the conditions $T_A,G_{eff}>0$, which can not be met for the analysis here.  We found an impossibility to define positive entropy and Misner-Sharp energy in any interval of physical time, as well as prevent ghost structure in the quantized theory \cite{starobinsky}. This suggests that the cosmological thermodynamic theory to $f(T,\mathcal {T})$ may not be well established fundamentally.
\par 
{\bf Our perspectives are listed as follows.
\par 
We use here the reconstruction method to cosmological models, which can then be examined reconstruction to other known models like Holographic and Chaplygin Gas for example. These new approaches can not modify the main result of the energy-momentum tensor conservation, the action is linear in the trace $\mathcal{T}$, more should cooperate with the acceleration of the universe at this stage. This will be covered in detail in a next work.
\par 
Another possibility approach is when we consider the space-time connection, regardless of the set of tetrads, then one of the dynamic fields. That would be a new version of $f(T,\mathcal{T})$ Cosmology in the Palatini's formalism, which may reveal new results in the reconstruction of the $\Lambda$CDM model. This approach can also reveal a different constraint to the conservation of energy tensor-momentum. Using the formalism reconstruction from the LQC model, as seen in \cite{LQC}, we can generalize to a non-minimum  coupling with the matter. We see in \cite{LQC} the reconstruction by an inverse method considering also $\rho\equiv\rho(\mathcal{R})$ or yet $\mathcal{R}\equiv \mathcal{R}(\mathcal{T})$, what is obtained in terms of the torsion scalar on \eqref{Tcal}. Could still think of the non-conservation of momentum-energy tensor as a consequence of quantum gravity anomaly \cite{Bertlmann}, then giving the possibility of action contain non-linear terms on the trace $\mathcal{T}$. This should be addressed in future work. The non conservation may also be arising from quantum effects such as creating particles in the expansion of the universe, which also occurs in Rastall's Cosmology \cite{Batista}.}
\vspace{1cm}

{\bf Acknowledgement}: Ednaldo L. B. Junior thanks PPGF-UFPA for the
hospitality during the realization of this work and also CAPES for financial support. Manuel E. Rodrigues  
thanks UFPA, Edital 04/2014 PROPESP, and CNPq, Edital MCTI/CNPQ/Universal 14/2014,  for partial financial support.
 Ines G. Salako thanks  IMSP  for hospitality during the elaboration of this work.


%

\end{document}